\documentclass{elsart}
\usepackage{epsfig}
\begin{document}

\begin{frontmatter}
\title{No end of tricks: electrons in the fractional quantum Hall regime}
\author{A.H. MacDonald}
\address{Department of Physics, Indiana University,
Bloomington, IN 47405, USA.}
\maketitle

\begin{abstract}
In the strong magnetic field
fractional quantum Hall regime, electrons in a two-dimensional
electron system are confined to their lowest Landau level.
Because of the macroscopic Landau level degeneracy
nearly all physical properties at low energies and
temperatures are then entirely determined by electron-electron
interactions.  The properties of these non-Fermi-liquid
electronic systems continue to surprise.  In this
article we briefly survey some recent advances in
the physics of the fractional Hall regime.
\end{abstract}
\end{frontmatter}

\section{Introduction}

Electrons in metals and doped semiconductors typically
form Fermi liquid states with qualitative physical properties
identical to those of a non-interacting
fermion gas state.  Exceptions can occur
when the electrons are strongly correlated.
The degree of correlation in typical electronic systems
depends on the ratio of the interaction strength to
the electronic Fermi energy or bandwidth.
Consequently interactions and correlations frequently play an essential
role, for example in giving rise to insulating rather than
metallic behavior, in the physics of narrow-band systems.

The role of interactions is particularly important for
two-dimensional electron systems (2DES) in a strong magnetic field.
In a magnetic field classical electrons move in circular cyclotron orbits
with angular frequency $\omega_{c} = e B / m c$.  (Here $e$ is the
magnitude of the electron charge, $B$ is the magnetic field strength,
$m$ is the electron mass.)  When the cyclotron motion is treated quantum
mechanically \cite{leshouches}, allowed kinetic energy values for a
single electron in a magnetic field are quantized;
 $\epsilon_{n} = \hbar \omega_{c} (n + 1/2)$, where $n$ is an
integer.  The kinetic energy of a cyclotron orbit is independent
of the location of the center of the orbit and, correspondingly,
there are many allowed single-particle
states with a given quantized kinetic energy.
(The Landau level degeneracy $N_{\phi} = A B / \Phi_{0}$ where
$A$ is the area of the system and $\Phi_{0} = hc/e $ is the
magnetic flux quantum. $\nu \equiv N/N_{\phi}$ is the Landau
level filling factor.)
A macroscopically degenerate set of single-particle states
with a common kinetic energy eigenvalue is known as a Landau level.
The fractional quantum Hall regime
occurs when the number of electrons is smaller than
the number of single-particle
states available in the lowest kinetic energy Landau level.
In this regime the number of {\it many-particle} states
which are degenerate in the absence of interactions diverges
exponentially with system size.  The Landau level in
most senses plays the role of a band with zero energetic width.
The perturbative treatment of interactions,
which we use to understand Fermi liquid behavior in
typical electronic systems, fails utterly.

All the physics discussed in this survey occurs in
the strong magnetic field limit in
systems for which electron-electron interactions
dominate over electron-disorder interactions \cite{iqhealso}.
The realization of such systems
required the development of modulation doping techniques
for the creation of high-mobility two-dimensional
layers in semiconductors.
Since these systems have become available,
experiments have revealed
a startingly rich variety of correlation effects, many of
which are still only partially understood.
The fractional quantum Hall regime has proved to be a wonderful
laboratory in which a wide variety of strong correlation
behaviors, most notably the fractional quantum Hall
effect \cite{tsui,laughlin} itself,
occur in well characterized experimental systems.  This
article discusses several phenomena which have
received recent attention \cite{qhereviews}.  It is
directed primarily toward researchers in other subfields
of solid state physics.

This article is organized as follows.  In Section II we present
a capsule summary of some essential features of fractional
quantum Hall physics which will be helpful in appreciating
subsequent sections.  One consequence of
interactions which can occur in electronic systems and which
\textit{does} occur in the quantum Hall regime is ferromagnetism.
In Section III we summarize recent work on the
unique properties of quantum Hall ferromagnets.
Interactions in the quantum Hall
regime can also lead to energy gaps at the Fermi energy.
When the chemical potential
lies in one of these gaps there are always states localized
at the edge of the system.  Recent work on the unusual physics of these
effectively one-dimensional `edge-state' systems is summarized in
Section IV.  Interaction and
correlation effects in an electronic systems often lead to
suppressed tunneling near zero bias.  These effects are
very strong in the fractional Hall regime and are not
yet completely understood.  Recent work on this topic is
outlined in Section V.   One important topic,
composite fermion behavior of electrons in the quantum Hall regime,
has been covered \cite{stormerssc} in a previous \textit{Highlights} issue
of \textit{Solid State Communications} and will be omitted here.
(That report is partially updated elsewhere in this volume
\cite{shayegan}.) We conclude in Section VI with some brief closing
remarks.

\section{Fractional Quantum Hall Physics in a Nutshell}

The perfect degeneracy of a Landau level appears to make
the kinetic energy portion of the Hamiltonian irrelevant
for fractional quantum Hall physics.  This appearance is
deceiving however, since the restriction of the Hilbert
space to single-particle orbitals with the minimum
kinetic energy constrains correlations contained in the
many-particle wavefunction \cite{anfunc}.  Without this constraint
the electrons would behave classically and, in their ground
state, would form a triangular lattice Wigner crystal.
Indeed electron crystallization was expected before experimental
studies of the fractional Hall regime
became possible.  (The Wigner crystal state
does in fact occur when the Landau level
is nearly empty.)   An important property
of the constrained Hilbert space is that it contains
only one relative motion state of a pair of electrons
for each relative angular momentum \cite{onlypositive}.
The Haldane pseudopotential \cite{haldane} $V_{M}$ is the
interaction energy of a pair of electrons with relative angular momentum $M$;
$\{V_{M}\}$ completely specify the projection of the Hamiltonian on to
the lowest Landau level.  Low energy states in the many-particle Hilbert space
are those for which pairs of electrons are unlikely to be in
the small $M$ relative angular momentum states in which
the repulsive interactions are stronger.

The quantum Hall effect is a transport anomaly which occurs
in the fractional quantum Hall regime.  It is characterized
by dissipationless current flow in the limit of zero
temperature.  The occurrence of the quantum Hall effect
can be related to the occurrence of an anomaly in
a thermodynamic property, the compressibility.
The compressibility of a system of interacting particles is
proportional to the derivative of the chemical potential with respect to
density. It can happen that at zero temperature the chemical potential has
a discontinuity at a density
$n^{*}$: the energy to add a particle to the system ($\mu^{+}$) differs,
at this density, from the energy to remove a particle from the system
($\mu^{-}$). The system is then said to be incompressible.
In an incompressible system a finite energy is required to create unbound
positive and negative charges that are capable of carrying current
through the bulk.  For this reason incompressible systems are as a rule
insulating at zero temperature.  Paradoxically,
incompressibility is precisely the condition required for
the quantum Hall effect to occur.  The twist is that in the case of the
quantum Hall effect, the density $n^{*}$ at which the incompressibility
occurs must depend on magnetic field.

The relationship between incompressibility and the transport
anomalies that give the quantum Hall effect its name
has been explained from several connected
points of view \cite{commsqhe} which are all consistent with
the following conclusions.  In the limit of zero temperature
the dissipative conductivity ($\sigma_{xx}$) at density $n^{*}$ vanishes and
the Hall conductivity approaches the value
\begin{equation}
\sigma_{xy} = (e^{2}/h) \Phi_{0} (\partial n^{*} / \partial B).
\label{eq:qhcond}
\end{equation}
For incompressibilities that occur at fixed Landau level filling
factor $\nu$, $\sigma_{xy} = (e^{2}/h) \nu$.
We will see in the following
sections that the magnetic field dependence
of $n^{*}$ is directly responsible for a number of unusual properties
at nearby densities.
(The quantum Hall effect also occurs at weaker magnetic
fields when many Landau levels are occupied.  The
gap in this case is usually due to kinetic energy quantization
and, since $\nu$ is then always an integer, the transport anomaly
is referred to
as the integer quantum Hall effect.  This survey is limited to
the fractional Hall regime or, taking account of the electronic
spin degree of freedom, to the filling factor range $\nu <2$.)

The largest of the chemical potential gaps
responsible for the fractional quantum Hall effect occurs
when the Landau level filling factor $\nu = 1/3$.
The origin of this gap is readily understood.  It turns out
that the largest Landau level filling factor at which it is
possible \cite{leshouches} to form many-body states
which completely avoid pairs with relative angular momentum
$M=0$ and $M=1$, and hence avoid the most repulsive interactions, is
$\nu = 1/3$.  The many-body states which satisfy this constraint
are the ones discovered by Laughlin \cite{laughlin} in his
seminal theoretical work on this
topic.  When $\nu =1/3$, an added electron must form a pair state
with relative angular momentum $M=1$ and it interacts more strongly
with the electron fluid than a removed electron.  The resulting
chemical potential gap occurs at fixed Landau level filling
factor and therefore at a density which is proportional to
magnetic field strength.  From transport experiments it is
known that chemical potential gaps occur at other filling with
the largest gaps occurring when $\nu = n / (2n +1)$ or
$\nu = (n+1)/(2 n +1)$.  The composite fermion
picture \cite{jaincf,hlr,stormerssc} provides
an intuitively appealing way of understanding why gaps occur
at these particular filling factors.
The existence of chemical potential gaps at certain filling factors
is the non-perturbative interaction effect in the fractional Hall regime
which is most dramatically manifested in transport experiments.
It is not by any means the only one.   It seems that every
new class of experiments and every theoretical advance leads to
unexpected conclusions.  In the following sections we discuss
some recent surprises.

\section{Quantum Hall Ferromagnets}

At first site the phrase `quantum Hall ferromagnet' appears
to be an oxymoron since ferromagnetism refers to spontaneous
magnetization in the absence of an external magnetic field
while the quantum Hall effect occurs in a 2DES
in the extreme strong magnetic field limit.
To understand why the terminology is sensible
it is necessary to consider the relevant energy scales for
the case of the semiconductors in which 2DESs are realized.
For a free-electron system in a magnetic field, the Zeeman
splitting of spin-levels $ g \mu_{B} B $ and the Landau
level separation $\hbar \omega_{c}$ are identical, apart from
small relativistic corrections. Electrons in states near the
conduction band minimum of a semiconductor behave like
free electrons except that band effects
renormalize the electron mass $m^{*}$ and the g-factor.
In the case of the GaAs system, where the quantum Hall effect
is most often studied, band effects increase the Landau
level separation by a factor of $\sim 20$ and reduce the
Zeeman splitting by a factor of $\sim 4$. As a result for
typical experimental situations, the Landau level separation
(in temperature units) is $\approx 150 {\rm K}$, and the
characteristic scale for electron-electron interactions
is $\approx 100 {\rm K}$ while the Zeeman splitting is only
$\approx 2 {\rm K}$.   We call a system a quantum Hall ferromagnet
if the electronic spins in an incompressible ground state
align in the absence of Zeeman coupling.
The small typical Zeeman coupling plays an important
role in determining measurable properties,
but it is still useful to treat the system as a ferromagnet
in the presence of a small symmetry breaking field.

If we regarded a partially filled Landau level as a giant
open-shelled atom the familiar Hund's rules from atomic physics
would suggest that, in order to maximize the magnitude of its
exchange energy, the ground state should have its spins aligned
to the maximum degree consistent with the Pauli exclusion principle.
In fact, a Hartree-Fock approximation would predict
a ferromagnetic ground state for electrons in a partially
filled Landau level at nearly any value of $\nu$.
The Hartree-Fock approximation, in which
many-electron states are approximated by single Slater
determinants, often provides a simple and qualitatively correct
picture of itinerant electron ferromagnetism at $T=0$.
In the fractional Hall regime, however, the truth is not so simple.
For example, at certain filling factors it is
known \cite{singlet} that the interaction energy is minimized in a $S=0$
state.   There are nevertheless some filling factors at which
the Hunds rule picture of the ground state is, perhaps coincidentally,
correct.  We will focus our attention here on the
$\nu = 1$ case which has been most extensively studied
theoretically and experimentally.  In this case the ground state is a
$S = N/2 $ spin-multiplet.  The states in this multiplet
are the only ones in the many-particle Hilbert space for which
pairs of electrons with angular momentum $M=0$ are completely
avoided and that is why they minimize the interaction energy.
An infinitesimal Zeeman coupling picks one state out of this
multiplet and completely aligns all the electronic spins.
The ground state has all majority spin orbitals in the lowest Landau
level occupied and all minority spin orbitals empty.
The same physics which is responsible for the ferromagnetic
ground state at $\nu =1$ gives rise to a chemical potential
gap when the density crosses $n^{*} = B/ \Phi_{0}$, as discussed
above.  The chemical potential gap makes a quantum Hall ferromagnet
similar to an insulating ferromagnet at low temperatures and
energies, although its itinerant character becomes important
at higher temperatures and energies.

Quantum Hall ferromagnets are isotropic ferromagnets, \textit{i.e.}
the ground state moment can point in any direction.   They are
like ferromagnets in a Heisenberg spin model and unlike those
in $XY$ where the moment orientation is confined to a plane
or Ising models where the moment must orient along a particular
axis.  On long length scales, the low-energy
states of all ferromagnets can be specified by a unit vector
field, $\hat m (\vec r)$, which specifies the (coarse grained)
local orientation of the ordered moment.
Two-dimensional isotropic ferromagnets have
excitations in which the moment field $\hat m(\vec r)$ has
a topologically non-trivial character.
These excitations are known as Skyrmions \cite{originalrefs,rajaraman}
and they are similar to the familiar vortex excitations
of $XY$ ferromagnets, except that their excitation energies are
finite.   When a small Zeeman field is present it aligns the moment
orientation far from the Skyrmion center.  The moment orientation
at the center of a Skrymion is then opposed to the Zeeman field.
The orientation of the perpendicular component of the moment
,which vanishes in magnitude at both zero and large distances from the
Skyrmion center, behaves like the orientation of the moment
near a vortex in an $XY$ magnet.  It rotates once around
the Zeeman axis along closed paths which enclose the Skyrmion center.
These Skyrmions are invariant under global rotation of all spins
around the Zeeman axis so that Skrymion states occur in families
whose members are labeled by, for example, the orientation of the
perpendicular component of the moment along a ray with a particular
orientation with respect to the Skyrmion center.  This degree
of freedom of a Skyrmion is important in understanding states where
the Skyrmions become dense, as discussed below.

\begin{figure}
\epsfysize=8cm
\centerline{\epsfbox{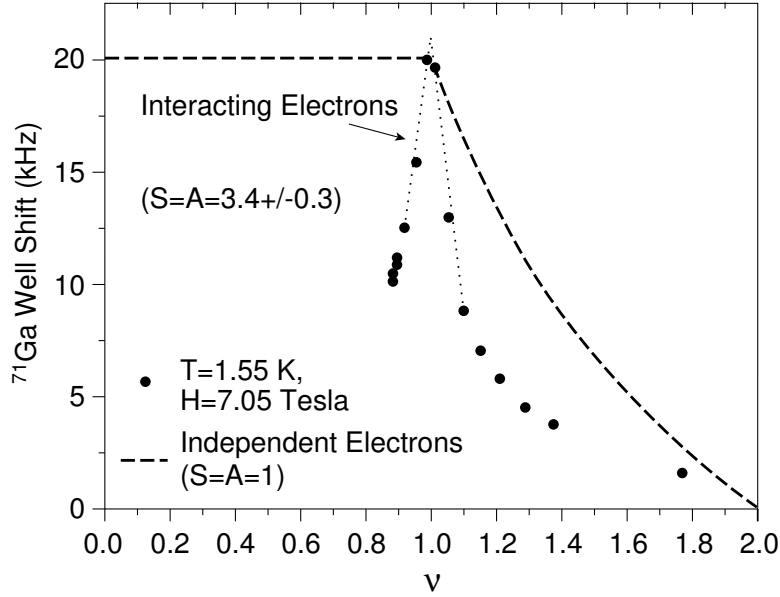}}
\vspace{6pt}
\caption[]{Knight shift measurements by Barrett \textit{et al.}
for a two-dimensional electron gas near filling
factor $\nu =1$.  The Knight shift is proportional
to the spin-polarization of the electron system which is, as
explained in the text,
expected to be fully polarized for $\nu =1$.
The experiment is consistent with four reversed
spins for each particle added to a full Landau level and
three reversed spins for each particle removed from a full
Landau level.  The dashed line
in this figure shows the Hunds rule prediction for the spin-polarization,
the maximum value consistent with the Pauli exclusion principle.
Full spin-polarization is possible when the filling factor $\nu < 1$;
at larger filling factors minority spin levels must be occupied.
The dependence of spin-polarization on
filling factor expected for non-interacting electrons would
be correct for interacting electrons if the Hartree-Fock
approximation predictions were correct.
(After Ref.~\cite{barrett}).}
\label{fig:one}
\end{figure}

Skyrmion excitations of this type are present in any two-dimensional
isotropic magnetic system.  Skyrmions in the quantum Hall effect
have the unique property that they carry charge \cite{leekane} $e$,
and this is responsible for their enormous impact on physical
properties.  In a pioneering paper on quantum Hall ferromagnets
Sondhi \textit{et al.} pointed out \cite{sondhi} that in typical
circumstances, Skyrmions should be the lowest energy charged excitations of
quantum Hall ferromagnets at $\nu =1$.  This work provided a
satisfying explanation for earlier numerical work by Rezayi
\cite{ednumerical} which hinted tantalizingly at unusual spin magnetism
near $\nu =1$. In the limit of large Skyrmion sizes, the energy of a
Skyrmion can be calculated from a generalized non-linear sigma
model \cite{originalrefs,rajaraman,sondhi}.
Because of the spin-reversal at the center, any
finite Zeeman coupling favors small Skyrmions.  This tendency
is countered predominantly by the Coulomb charging energy
of a Skyrmion.  For Zeeman coupling strengths typical of
experimental systems, an accurate calculation of
the energy of a Skyrmion and of the number of reversed spins each contains
near its center requires a microscopic calculation.  The
required calculation was completed \cite{fertigskyrmion} by Fertig
\textit{et al.} who predicted approximately three reversed spins per
quasiparticle near $\nu =1$ and suggested that the presence of Skyrmions
would be manifested most clearly in measurements of the
filling factor dependence of the ground state spin-polarization.
Coincidentally, Barrett and Tycko \cite{barrett} developed an optical
pumping technique to achieve the first NMR studies of the two-dimensional
electron gas.  Their results for the low-temperature
Knight shift, which is proportional to the ground state spin
polarization are shown in Figure ~\ref{fig:one}.

These results can be understood as follows.
For $N = N_{\phi}$ the ground state has $S_{z}= S =N/2$
as discussed above. For $N= N_{\phi} \pm 1$, the ground state
contains a single charged Skyrmion. The Skyrmion can be introduced
by changing the total electron number or, in what is the typical
experimental situation, by changing the magnetic field strength
and hence $N_{\phi}$.   In a quantum description \cite{hcmskyrm,nayak}
the number of reversed spins per skyrmion is quantized so
that, when Skyrmion-Skyrmion interactions can be
neglected, we expect that the component of the total spin
along the direction of the Zeeman field is
\begin{equation}
S_{z} = N/2 - (K + \theta) |N - N_{\phi}|
\label{eq:szqn}
\end{equation}
Here, in reflection of a particle-hole symmetry which exists in the
fractional Hall regime \cite{phsym}, $\theta = 1$ for $N > N_{\phi}$
and $\theta =0$ for $N < N_{\phi}$.
$|N - N_{\phi}|$ is the number of Skyrmions or antiskyrmions present
in the system.   The Knight shift measurements \cite{barrett} of Barrett
{\it et al.} are consistent with $K=3$ and with the Hartree-Fock
calculations \cite{fertigskyrmion} of Fertig \textit{et al.}.
For non-interacting electrons, or with interactions treated in the
Hartree-Fock approximation, $K=0$ so that $S_{z}$ always has the
maximum value allowed by the Pauli exclusion principle.
There seems to be little doubt that the elementary charged
excitations of quantum Hall ferromagnets are Skyrmion-like
objects that carry large spin quantum numbers.
Recent transport \cite{andy} and optical \cite{goldberg95}
experiments add additional support to this conclusion.

For large enough $| N - N_{\phi}| $ the Skyrmion-like
objects will eventually interact strongly.
When the density of Skyrmions is low and the temperature
is low, Skyrmions are expected to form a triangular lattice crystal
similar to the Wigner crystal state formed by
electrons in the limit of very strong magnetic fields.
The large range of filling
factors over which $S_{z}$ falls linearly with $\nu -1$ indicates
that Skyrmions tolerate crowding with amazing alacrity.
This property has been explained \cite{sklatpaper} by Brey \textit{et al.}
who have proposed that at higher densities Skyrmions will
form a square lattice.  The basis of this proposal, supported by
detailed microscopic calculations, is the observations that
neighboring Skyrmions repel each other less strongly when
they have opposite orientations.   Dense Skyrmions will
thus tend to form a bipartite lattice, most likely square
, with long range order in the Skyrmion orientations.
At a finite temperature
this state has both quasi-long-range translational and
magnetic order, leaving open the possibility of one or
two continuous Kosterlitz-Thouless or first order phase
transitions at finite temperature.  Recent anomalies \cite{bayot}
seen in heat capacity studies of 2DESs in the quantum Hall regime
near $\nu=1$ may reflect such a phase transition.

\begin{figure}
\epsfysize=8cm
\centerline{\epsfbox{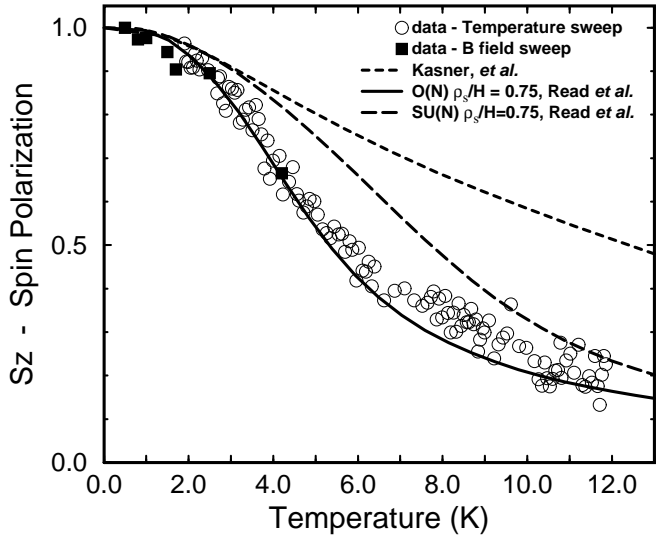}}
\vspace{6pt}
\caption[]{Temperature dependence of the spin-polarization
as determined from polarization dependent optical absorption
measurements. The open circles and closed squares are experimental
data.   The solid and long-dashed lines show two different
approximate theoretical results calculated by Read and
Sachdev in Ref.~\protect \cite{readqhf}
using a field-theoretical model with a realistically chosen spin-stiffness
parameter.  The short-dashed line is a theoretical
result obtained in Ref.~\protect \cite{kasner} using
an approximation which accounts for the
itinerant character of the two-dimensional electrons,
important at higher temperatures.(After Ref.~\cite{goldberg96}.)
}
\label{fig:two}
\end{figure}

The finite temperature properties of quantum Hall ferromagnets
are equally interesting.   The simplest situation, which occurs when
$\nu =1$ and no Skyrmions are present in the ground state,
has received the greatest attention to date.  In Figure ~\ref{fig:two}
we show measurements of the temperature dependence
of the magnetization at $\nu =1$ by Goldberg\cite{goldberg96}
and collaborators as determined using polarization dependent optical absorption
measurements.  Similar results were obtained earlier in the 
NMR studies of Barrett and Tycko \cite{barrett}.  
According to the Mermin-Wagner theorem, long range
spontaneous magnetic order is impossible in two-dimensional
systems at any finite temperature.  The thermal suppression
of the magnetization at very weak Zeeman coupling is
expected to exhibit quantum critical behavior.
Most of the thermal magnetization suppression occurs at
temperatures low enough that the system is
completely characterized by its spin stiffness energy, $\rho_{s}$.
In Figure ~\ref{fig:two}
the measured magnetization is compared with theoretical
approximations to the temperature dependence of magnetization
calculated by Read and Sachdev \cite{readqhf} using the appropriate
continuum quantum field theory model.
The field theory description used for the long-wavelength
physics is appropriate for any two-dimensional ferromagnet with
a magnetic correlation length which is long compared to microscopic
lengths and which has no low energy excitations other
than those associated with slow smooth magnetic fluctuations.
The validity of the continuum quantum field theory model at low temperatures
in the itinerant two-dimensional electron system
is a consequence of the chemical potential gap for particle-hole excitations
at $\nu =1$.  In the limit where the field theory applies
the magnetization depends only on the independent ratios of $k_{B} T$,
$\rho_{s}$ and the Zeeman energy.  

At higher temperatures the magnetization curves
are no longer universal and the itinerant nature of
the underlying electronic degrees of freedom becomes important,
for example in screening the Coulomb interactions responsible
for the spin stiffness.   The need for a simultaneous treatment of collective
spin and constituent fermionic degrees of freedom produces a fundamental
difficulty.  Lack of progress on this front has blocked the development
of a completely satisfactory theory of itinerant electron magnetism.
Quantum Hall ferromagnets present
a, possibly especially simple, system in which to compare
theoretical treatments of itinerant electron ferromagnets with
experiment.  The $\nu =1$ quantum hall ferromagnet
is in many senses similar to a strong ferromagnetic ground
state in a single-band itinerant electron system.
In Figure ~\ref{fig:two} we show theoretical
results for the magnetization obtained in an approximation\cite{kasner}
which does not account for the spin-wave interactions important
in the quantum critical regime but does account for the
emergence, at high temperatures,
of the underlaying fermionic degrees of freedom
upon which the spin-waves are built.  Evidently this
theory overestimates the magnetization at high temperatures.  Possible
sources of this discrepancy
include an inadequate treatment of screening, which
certainly reduces the spin stiffness at higher temperatures, and the neglect
of the spin-wave interactions important at lower temperatures and
Zeeman energies and included in the field theoretical approximations.
Progress in identifying a major culprit and success in making
appropriate improvements to the theory could have important
implications for theories of itinerant electron magnetism.
Understanding the temperature dependence at nearby filling
factors \cite{barrett} presents an especially interesting 
challenge to theory. 

Physics closely related to that discussed above occurs
in systems with two 2D layers close together, even at stronger
fields where the electronic system is always fully
spin polarized.\cite{dassarmabook}
The role of the the spin quantum number is assumed by
the quantum number which specifies the layer in which an
electron resides.   The role of the Zeeman energy
is assumed by the tunneling gap between symmetric and
antisymmetric single-particle states.  There are, however,
a number of important differences between the spin and
double-layer cases.  It turns out that, when the double-layer
system is described using a pseudospin language, the
effective magnetic interaction in the double-layer case
is anisotropic leading to an easy-plane ordered state.  The broken symmetry
in this ordered state results in inter-layer phase coherence
in the absence of interlayer tunneling.  The
topologically and electrically charged excitations
gradually deform with increasing anisotropy
from skyrmions to vortex-antivortex pairs.  The physics
of double-layer systems is further enriched by the
possibility of disturbing the system with a magnetic
field parallel to the electronic planes.  In the
pseudospin description such a field gives rise, in a
system with non-zero symmetric-antisymmetric splitting, to
an effective Zeeman field which rotates as a function of
the perpendicular planar coordinate and at a critical
in-plane field strength causes
a phase transition.  The mathematical description of this
phase transition is identical to the description of
commensurate-incommensurate phase transitions.  This
phase transition was first discovered in measurements\cite{sheena} of
the transport properties of double-layer systems.  This experimental
work actually predates the experimental studies of
quantum Hall ferromagnets surveyed in the preceding paragraphs.
Further progress in unraveling the richer physics of double-layer
quantum Hall ferromagnets has been hampered, however, by
slower headway in the development of pertinent experimental
probes.  It appears that progress has accelerated recently,
however, with the demonstration that optical probes can
measure the pseudospin moment \cite{pellegrini}. 

\section{Edge Excitations of an Incompressible Quantum Hall Fluid}

\begin{figure}
\epsfysize=9cm
\centerline{\epsfbox{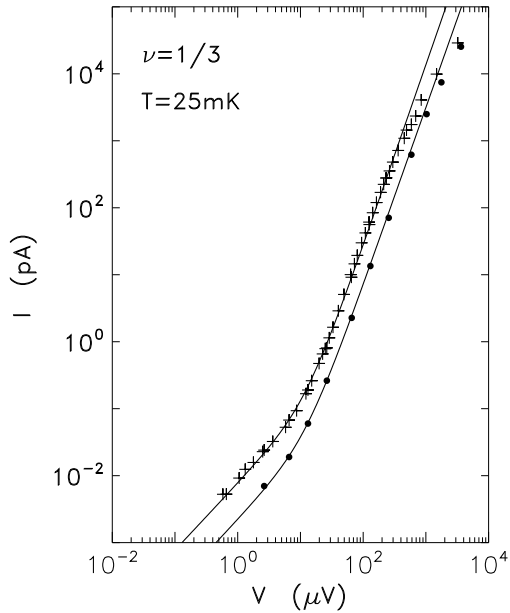}}
\vspace{6pt}
\caption[]{
Log-log plot of the current-voltage ($I-V$) characteristic
for tunneling from bulk-doped n+ $GaAs$ into the edge of
a two-dimensional electron system with $\nu=1/3$.
Results are shown for two different samples with
different densities for which $\nu =1/3$ occurs at
$B=13.4T$ (crosses), and $B=10.8T$ (solid circles).  The solid curves
represent fits to the universal form predicted theoretically
by Kane and Fisher in Ref.~\cite{KF}.
(After Ref.~\cite{changref}.)}
\label{fig:three}
\end{figure}

When the chemical potential lies in a gap a finite fractional
Hall system has no gapless excitations \cite{spincaveat}
in the bulk but it can have gapless excitations localized
at its edge, and in fact must \cite{mylecturenotes}
have such excitations whenever the Hall conductivity is non-zero.
In Figure ~\ref{fig:three}
we show some recent experimental
results obtained by Chang \cite{changref} on the tunneling
density-of-states for a fractional Hall system at filling
factor $\nu = 1/3$.  What is plotted here is the measured tunneling
current between an effectively metallic contact and the edge of the quantum
Hall system.  Assuming that there are no anomalies in the
density-of-states of the contact, the current
is proportional to the tunneling density-of-states
integrated over an energy interval $eV$ starting from the chemical
potential.  A finite tunneling density-of-states would lead to a
tunneling current $I \propto V$ and a tunneling
conductance which approaches a constant at low-temperatures.
Instead Chang's experiments show a tunneling current $I
\propto V^{\alpha}$ (in the low temperature limit) and
a tunneling conductance proportional to $T^{\alpha-1}$
where $\alpha$ is $\simeq 2.7$.  These results are
in substantial agreement with earlier theoretical predictions
by Wen \cite{wenedge} and Kane and Fisher \cite{KF}.
Evidently the density-of-states is strongly suppressed at
energies near the Fermi energy in the $\nu =1/3$ case.

The edge of a two-dimensional fractional
Hall system constitutes a one-dimensional electron system.
At energies well below the chemical potential gap,
the incompressible fractional Hall system can be
described using a chiral Luttinger liquid picture \cite{wenreviews}
which is an adaptation of the Luttinger liquid theory \cite{haldanell}
for low-energy properties of one-dimensional electron systems.
Low-energy excitations of the electronic system are expected
to involve modulations of the edge which are slow on a microscopic
length scale.  The simplest version of the chiral Luttinger
liquid picture starts from the assumption that quantum states
at the edge are completely characterized by the one-dimensional
density [$n(x)$] obtained by integrating the change in the
two-dimensional density from its ground state form along a
coordinate perpendicular to the edge.
(We'll comment further on this assumption, which is not always
valid, later.)  Assuming translational invariance along the edge and
expanding around the ground state to describe low-energy excitations
gives
\begin{equation}
E[n] = E_{0} + \frac{1}{2} \int dx' \; \int dx\;
 \delta n(x') \alpha(|x - x'|) \delta n(x)
\label{eq:enlnr}
\end{equation}
where $E_{0}$ is the ground state density and $\alpha(x) $
is a non-universal function.  ($\alpha (x)$ must depend
at a minimum on the details of the external potential
which confines the electronic system to a finite area.)
Note that, for excitations which change the total electron density,
we have as a convenience chosen the zero of
energy at the chemical potential in dropping a term proportional
to $ \int dx \, \delta n(x)$.  When expressed in terms of
the Fourier components of
$n(x)$ Eq.~\ref{eq:enlnr} becomes
\begin{equation}
E = E_{0} + \frac{1}{2L} \sum_{q\neq 0} \alpha_{q}
n_{-q} n_{q}.
\label{eq:edgeeng}
\end{equation}
For the purposes of the discussion below we will assume that
$\alpha_{q}$ approaches a constant (to be denoted by $\alpha$)
in the $q \to 0$ limit
appropriate at low energies \cite{lrcaveat}.
The energy above can be used as an effective Hamiltonian for low-energy
long-wavelength excitations.  It is quantized by
imposing the following commutation relation on the density Fourier
components:
\begin{equation}
[n_{-q'},n_{q}] = \frac{\nu qL}{2\pi}\; \delta_{q,q'}.
\label{eq:commutator}
\end{equation}
This commutation relation is suggested by the Luttinger liquid
theory of the one-dimensional electron gas.  The innocent looking
factor of $\nu$ on its right hand side turns out to be
responsible for the suppressed tunneling density-of-states
when $\nu \ne 1$.  As we explain below, it is required in order
for the theory to imply the correct value of the quantized
Hall conductance.

Eq.~\ref{eq:commutator} and Eq.~\ref{eq:edgeeng} imply that
the quantized density-wave excitations at the edge are
equivalent to a system of non-interacting bosons.
The boson creation and annihilation operators are defined by
\begin{eqnarray}
a_{q} &=& \sqrt{\frac{2\pi}{\nu qL}}\; n_{-q}\\
a_{q}^{\dagger} &=& \sqrt{\frac{2\pi}{\nu qL}}\; n_{q}
\end{eqnarray}
where $q > 0$ so that all the density waves travel in the same
direction, hence the chiral designation for this Luttinger
liquid system.  The Hamiltonian is
\begin{equation}
H = \sum_{q>0} \hbar\; vq\; a_{q}^{\dagger} a_{q}^{\phantom{\dagger}}
\label{eq:hamiltonian}
\end{equation}
so that the velocity of the edge waves is
\begin{equation}
v = \frac{\alpha \nu }{2\pi\hbar} = \frac{\nu}{2\pi L \hbar}\;
\frac{d^{2}E_{0}}{dn^{2}} = \frac{\nu}{2\pi\hbar}\;
\frac{d\mu}{dn}
\label{eq:velocity}
\end{equation}

The justification of the chiral Luttinger liquid theory in
the case of the fractional quantum Hall effect ($\nu \ne 1$)
is not as systematic as in the one-dimensional electron gas case.
As explained above, Eq. ~\ref{eq:commutator} implies that
low-energy excitations are chiral bosons.  For $\nu =1/m$
where $m$ is an odd integer
the bosonization property can be established from microscopic theory
\cite{mdj}, provided that the external potential which defines the edge
of the system is sufficiently abrupt.  For incompressible states at
other values \cite{mdj,wenreviews} of $\nu$, and for smoother edges,
the edge state electronic structure is more complicated.
In general neither Eq. ~\ref{eq:edgeeng} nor Eq. ~\ref{eq:commutator}
applies and more elaborate theories need to be
developed \cite{wenedge,wenreviews,macdedge}.  When this simplest
version of the chiral bosonization
{\it does} apply, the expectation of dispersionless propagation of
low-energy edge density waves requires that the
commutator $[n_{-q'},n_{q}] $ be $\propto q \delta_{q,q'}$.
The constant of proportionality can then be determined by
requiring that the rate of change of the equilibrium edge current with
chemical potential be $ e \nu / h$.   It can be shown that
this is equivalent to requiring that the theory reproduce the
correct quantized value of the Hall conductivity.
The change in equilibrium edge current is related to the change
in equilibrium density by
\begin{equation}
\delta I = ev\delta n
\end{equation}
When the chemical potential for the single edge system is shifted
slightly from its reference value (which we chose to be zero) the energy
is given by
\begin{equation}
E[n] = E_{0} + \mu\delta n + \alpha\frac{(\delta n)^{2}}{2}
\label{eq:nov13a}
\end{equation}
Minimizing with respect to $\delta n$ we find that
\begin{equation}
\delta n = \frac{\delta \mu}{\alpha}
\label{eq:nov13b}
\end{equation}
so that
\begin{equation}
\frac{\delta I}{\delta\mu} = \frac{ e v }{\alpha}
\label{eq:nov13c}
\end{equation}
In order for this to be consistent with the quantum Hall effect [$ \delta
I = (e \nu /h) \delta \mu $] our theory must yield a edge phonon
velocity given by
\begin{equation}
 v = \frac{\alpha}{2 \pi \hbar} \cdot \nu.
\label{eq:nov13d}
\end{equation}
The factor of $\nu$ appearing in this equation occurred in
Eq.~(\ref{eq:velocity}) because of the factor of $\nu$ in
Eq.~(\ref{eq:commutator}).  This factor is required if the theory
is to reproduce the correct value of the quantized Hall conductance.

In order to calculate the tunneling density-of-states we need to
express the many-particle state produced when an electron
is instantaneously added at the edge of
a two-dimensional electron system in an incompressible
state as a state of the bosonic edge wave system.
The relationship between electron and boson operators is
well-known from Luttinger liquid theory and
is established by requiring the exact identity
\begin{equation}
[n (x),\hat{\psi}^{\dagger}(x')] = \delta (x-x')\;
\hat{\psi}^{\dagger}(x')\;
\label{eq:creation}
\end{equation}
to be reproduced by the effective low-energy theory.  Here
$\hat{\psi}^{\dagger}(x')$ is the operator that creates an
electron at position $x'$ along the edge.  This equation simply
requires the electron charge density to increase by the required amount
when an electron is added to the system.
An elementary calculation \cite{wenedge,wenreviews,mahanll} shows that
in order to satisfy Eq.~(\ref{eq:creation}), the field operator must be
given by
\begin{equation}
\hat{\psi}^{\dagger}(x) = \sqrt{z}  e^{i \, \nu^{-1} \, \phi (x)}
\label{eq:nov13a3}
\end{equation}
where $d \phi(x) / dx = n(x)$ and $z$ is a constant that cannot be
determined by the theory.  The factor of $\nu^{-1}$ in the argument of
the exponential of Eq.~(\ref{eq:nov13a3}) is required because of the
factor of $\nu$ in the commutator of density Fourier components that in
turn was required to make the theory consistent with the fractional quantum
Hall effect. When the exponential is expanded the $k^{th}$ order terms
generate states with total boson occupation number
$k$ and are multiplied in the fractional case by the
factor $\nu^{-k}$; multi-phonon terms are increased
in relative importance.

Eq.~(\ref{eq:nov13a3}) has been carefully checked
numerically \cite{palacios}. The $\nu^{-1}$
factor leads to predictions of qualitative changes in a number of
properties of fractional edges. The quantity that is most directly
altered is the tunneling density-of-states.  Consider the
state created when an electron, localized on a magnetic length scale, is
added to the ground state at the edge of a $N-$ electron
system with $\nu =1/m$:
\begin{eqnarray}
\hat{\psi}^{\dagger}(0) |\Psi_{0}\rangle &\sim& \exp{\left(
-\sum_{n>0} \frac{a_{n}^{\dagger}}{\sqrt{n \nu}}\right)}
|\psi_{0}\rangle\nonumber\\
 &=& 1 + \frac{1\mbox{ phonon term}}{\nu^{1/2}} + \frac{2\mbox{ phonon
terms}}{\nu } + \ldots.
\end{eqnarray}
The tunneling density-of-states at zero temperature
is given by a sum over the ground and
excited states of the $N+1$ particle system:
\begin{equation}
A(\epsilon) = \sum_{n} \delta(E_{n} - E_{0} - \epsilon)
| \langle \Psi_{n} | \psi^{\dagger}(0) | \Psi_{0}  \rangle |^{2}
\label{eq:sfun}
\end{equation}
Because of the increased weighting of multiphonon states, which become
more numerous at energies farther from the chemical potential,
the spectral function is larger at larger $\epsilon -
\mu$ in the fractional case.  Explicit calculations \cite{wenedge,KF}
yield a spectral function at zero temperature that grows like
$(\epsilon-\mu)^{{\nu}^{-1}-1}$.  At finite temperature
the tunneling density-of-states saturates to a finite
value $\propto T^{{\nu}^{-1}-1}$.
The experimental results in Fig.~\ref{fig:three} are consistent
with the predicted energy dependence and with predictions for
the temperature dependence of the voltage at which saturation
is expected to occur.  These observations evidently confirm the physics
discussed in the preceding paragraphs.

It seems intuitively clear that the
spectral function should be small at low-energies in the fractional case
since the added electron will not share the very specific correlations
common to all the low-energy states. It is marvelous that simply
requiring the low-energy theory to be consistent with the fractional
quantum Hall effect leads to a very specific prediction for the way in
which this qualitative notion is manifested in the tunneling
density-of-states at the edge of a $\nu = 1/m$ incompressible state.
Similar predictions can be made for
the richer behavior \cite{KF,wenreviews} expected in tunneling
experiments at filling factors with more complicated edge
electronic structures,  for example at $\nu = 2/3$.
The experimental result shown in Fig.~\ref{fig:three} is
likely to be the harbinger of a period of fruitful interaction
between theoretical and experimental studies of fractional
Hall edges with important consequences for our understanding
of quasi one-dimensional electron systems more generally.

\section{Tunneling Density of States; Spectral Function}

\begin{figure}
\epsfysize=9cm
\centerline{\epsfbox{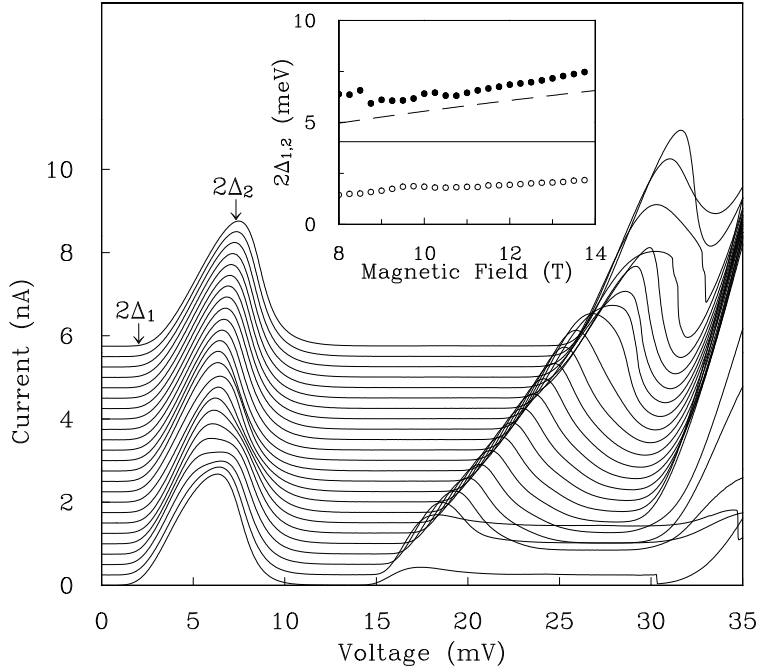}}
\vspace{6pt}
\caption[]{Low temperature tunneling I-V characteristics for bulk 2D to 2D
tunneling in the fractional Hall regime.  The traces are
at magnetic fields separated by $0.25$ Tesla and for
this sample cover a range of filling factors from
$\nu = 0.48$ to $\nu = 0. 83$.  The higher bias
potential peak is due to Landau level mixing while the lower
bias potential peak is due to tunneling within the lowest
Landau level.  The tunneling current is the convolution of
the tunneling densities of states in the two layers
at energies on opposite sides of the Fermi energy and
separated by $e V$.  This experiment shows that the tunneling current
is strongly suppressed near zero bias and hence that the
tunneling density-of-states is suppressed at energies near the
Fermi energy over a broad range of filling factors.  Features
in the tunneling data associated with particular
incompressible states, for example the one that occurs at $\nu = 2/3$, are
weak. The inset shows the onset and peak values of the intra
Landau level tunneling currents.
(After Eisenstein \textit{et al.} in Ref.~\cite{eisenstein}.)
}
\label{fig:four}
\end{figure}

In the preceding two sections we discussed experimental
achievements leading to the confirmation of phenomena
which had, at least in part, been anticipated theoretically.
The possibility of performing experiments will no doubt
in both cases lead to the further development and refinement of
theoretical ideas.  In this section we discuss experimental
observations which were not, as far as we are aware,
anticipated by theory.  These experiments were made possible
by the ability to make separate electrical contact to nearby
two-dimensional electron systems, enabling measurements
of the tunneling conductance from one bulk two-dimensional
electron system to another.  It was known from transport
studies that the bulk tunneling density-of-states must
be strictly zero near the Fermi energy at those discrete
set of filling factors for which the ground state is
incompressible and the transport anomalies of the fractional
quantum Hall effect occur.  Prior to the first experiments,
most experts would probably have
predicted a tunneling density-of-states which was depressed due
to interactions only at energies quite close to the Fermi energy
and only near incompressible filling factors.  What was
observed instead \cite{eisenstein,eisenstein95,pepper}
is shown in Fig.~\ref{fig:four}.
Clearly there is a strong suppression of the tunneling density-of-states (TDOS)
near the Fermi energy over a wide range of $\nu$ in the
fractional quantum Hall regime.  This suppression was also seen,
although less distinctly, in earlier 3D to 2D tunneling
experiments \cite{ashoori}.

This behavior for the tunneling conductance is in stark contrast
with the behavior in identical systems in the absence of a
strong magnetic field.  In that case, momentum conservation
and Fermi liquid theory,
which implies quasiparticle peaks in the electronic spectral
functions, conspire to
produce large \textit{peaks} in the tunneling current
\cite{expb0,theoryb0} at zero bias.  On a qualitative level the very
different results found for the non-Fermi liquid states of the strong
magnetic field limit are easy to understand on the basis of the physical
picture suggested in early experimental work \cite{eisenstein}.
As we have emphasized the ground state at any filling factor minimizes
the interaction energy, subject to the constraint that all
electrons lie in the lowest Landau level.  Without this constraint
the electrons would form a crystal.
The ground state can still be thought of as
a `quantum melted' crystal in which positional order is lost because
the fluctuations in electronic positions are too large.
In the tunneling experiment an electron is removed from the
ground state in one two-dimensional layer and added to
another layer.  The measurement tells us that the difference
between the energy for adding an electron and removing an
electron has a probability distribution which
is relatively sharply peaked at a finite value
and is extremely small values for energies close to zero.  This is what
would be expected if the ground state really were a
crystal.  In that case a suddenly removed electron would be
removed from lattice site, or close to a lattice site if
lattice vibrations were accounted for.  A suddenly added
electron could be inserted into the lattice at any point.
However even the energetically most advantageous interstitial sites
would would be less favorable than a lattice site.  The net energy cost
could be close to zero only if the lattice constant of the electron
crystal were adjusted to allow for the creation of an additional lattice
site for the added electron.   The probability of finding the
electronic system is such a convenient state is evidently very small
and therefore the tunneling current at very small bias voltages
must vanish very quickly.

There has been considerable theoretical interest in these
experimental results and a number of different
approaches \cite{othertheory,exactdiag} have been taken in an
effort to put these ideas on a firmer footing and make more
quantitative predictions which could be tested experimentally.
At present there is no completely satisfactory theory.
As is often the case, it is much easier to derive
rigorous results for moments of a spectral function than
for its full frequency dependence.   Following Haussmann \textit{et
al.} \cite{sumrule} we define an effective energy gap in
terms of the ratios of the first and zeroth moments of
the tunneling I-V curve as follows:
\begin{equation}
\Delta_{sr} = \frac{ \int_{0}^{\infty}\epsilon I(\epsilon ){\rm d}\epsilon}
{\int_{0}^{\infty} I(\epsilon ){\rm d}\epsilon}.
\label{eq:ivmom1}
\end{equation}
In the limit of sharp peaks in I-V curve, $\Delta_{sr}$ gives the
position of the peak.  It turns out that it is possible
to derive \cite{renn,sumrule} an exact relationship between $\Delta_{sr}$
and the ground state energy at a particular filling factor:
\begin{equation}
\Delta_{sr}
= \frac{2[\nu^{2} \epsilon_{0}(\nu =1) - \epsilon_{0}(\nu)]}{\nu (1 - \nu)}
\label{eq:srgap}
\end{equation}
In this equation, $\epsilon_{0}(\nu)$ is the ground state
energy per state in the Landau level
at filling factor $\nu$.  In the Hartree-Fock approximation
the ground state energy per single-particle state is
proportional to $\nu^{2}$, provided that translational symmetry
is not broken, so Eq.~\ref{eq:srgap} says that the gap
is proportional to difference between the ground state
energy in the Hartree-Fock approximation and the exact
ground state energy, \textit{i.e.}, to the amount by which the
ground state energy is lowered by correlations.  This result
is very much in accord with the physical picture given above.
The ground state energy per electron is fairly accurately
known and Eq.~\ref{eq:srgap} is in good agreement with experimental
results.  Any complete theory of bulk 2D-2D tunneling in the
fractional Hall regime will have to respect this exact identity.

It follows from Eq.~\ref{eq:srgap} that the filling factor dependence
of the ground state energy can be extracted from tunneling data,
in much the same manner as it has previously been extracted from
extrinsic photoluminescence data \cite{kukushkin}.  Cusps in the
filling factor dependence of the ground state energy correspond to
discontinuities in the chemical potential.  The cusp at filling
factor $\nu =2/3$ is presumably responsible for features seen in
the peak conductance voltage near $B = 9$ Tesla in
Fig.~\ref{fig:four}; it will be interesting to see how much
detailed information on the ground state energy will be extracted
from tunneling experiments in the future.

\section{Concluding Remarks}

This brief survey highlights some active areas of
electron-electron interaction physics in the fractional Hall regime.
It seems certain that future experimental and theoretical
activity will refine the present understanding
of all these topics and, if the past is any guide,
also lead to the discovery of phenomena which are not presently anticipated.
I hope that any blurriness the reader notices
in my snapshot of this \textit{tableau vivant}
inspires more intrigue than annoyance.
My views on the matters discussed here have been shaped by
discussions with members of the condensed matter theory group at Indiana
University, including M. Abolfath, S.M. Girvin, C. Hanna,
R. Haussmann, M. Kasner, S. Mitra, K. Moon, J.J. Palacios, D. Pfannkuche,
E. Sorensen, K. Tevosyan, K. Yang, and U. Z\"{u}licke.
Discussions with I. Aleiner, H. Baranger, S. Barrett, L. Brey, A. Chang,
R. Cote, A. Efros, J. Eisenstein, H. Fertig, M. Fisher, L. Glazman,
M. Johnson, C. Kane, L. Martin, J. Oaknin,
C. Tejedor, S.R.-E. Yang and X.-G. Wen are also gratefully
acknowledged.  My work is supported by the National Science Foundation
under grant DMR-9416906.


\begin{thebibliography}{}
\bibitem{onlypositive} Because of broken time reversal
symmetry in the strong magnetic field limit only
states with positive relative angular momentum occur.
\bibitem{spincaveat} An exception occurs when the ground
state is a quantum Hall ferromagnet and the Zeeman energy
is zero so that spin-wave excitations are gapless.
\end{thebibliography}
\end{document}